\documentclass[onecollarge,natbib]{svjour2}
\bibpunct{[}{]}{;}{n}{}{,} 
\smartqed  
\usepackage{graphicx}
\usepackage{bm}       
\usepackage{amsmath}  
\usepackage{amssymb}
\usepackage[utf8]{inputenc}

\journalname{Archive of Applied Mechanics}
\begin{document}

\title{ An algebraic model for the pion's valence-quark GPD 
}
\subtitle{A probe for a consistent extension beyond DGLAP region via Radon transform inversion}


\author{Nabil Chouika   \and C\'edric Mezrag \and Herv\'e Moutarde \and Jos\'e Rodr\'\i guez-Quintero
}


\institute{ J. Rodr\'\i guez-Quintero \at
               Dpt. F\'{\i}sica Aplicada, Fac. CCEE, Unversidad de Huelva, Huelva E-21071, Spain\\
              Tel.: +34-959-219787\\
              Fax: +34-959-219777\\
              \email{jose.rodriguez@dfaie.uhu.es}           
           \and
           N. Chouika \& H. Moutarde \at
              Centre de Saclay, IRFU/Service de Physique Nucl\'eaire, F-91191 Gif-sur-Yvette, France 
              \and 
            C. Mezrag \at
            Physics Division, Argonne National Laboratory, Argonne, Illinois 60439, USA.  
}

\date{Received: date / Accepted: date}

\maketitle

\begin{abstract}
  We briefly report on a recent computation, with the help of a fruitful algebraic model, sketching the pion valence dressed-quark generalized parton distribution. Then, preliminary, we introduce on a sensible procedure to get reliable results in both Dokshitzer-Gribov-Lipatov-Altarelli-Parisi (DGLAP) and Efremov-Radyushkin-Brodsky-Lepage (ERBL) kinematical regions, grounded on the GPD overlap representation and its parametrization of a Radon transform of the so-called double distribution (DD).

\end{abstract}

\section{Introduction}
\label{intro}

The Generalized Parton Distributions (GPDs), introduced independently by M\"uller \etal \cite{Mueller:1998fv}, Ji \cite{Ji:1996nm} and Radyushkin \cite{Radyushkin:1997ki}, are shown to be related to hadron form factors by sum rules, and to contain the usual Parton Distribution Functions (PDFs) as a limiting case.  They not only generalize the classical objects describing the static or dynamical content of hadrons; additionally, they also provide unique information about the structure of hadrons, including 3D imaging of its partonic components and access to the quark orbital angular momentum. GPDs have been the object of an intense theoretical and experimental activity ever since (see {\it e.g.} \cite{Tiburzi:2002tq,Theussl:2002xp,Broniowski:2003rp,Ji:2006ea,Broniowski:2007si,Frederico:2009fk}, for the pion case, or the more general reviews \cite{Ji:1998pc,Goeke:2001tz,Diehl:2003ny,Belitsky:2005qn,Boffi:2007yc,Guidal:2013rya} and references therein).

Most of the constraints that apply to GPDs are fulfilled when the GPD is written in a double distribution representation~\cite{Mueller:1998fv,Radyushkin:1998es,Radyushkin:1998bz}, which is plainly equivalent to expressing the GPD as a Radon transform \cite{Teryaev:2001qm}. In order to obtain insights into the nature of hadron GPDs, it has been common to model the Radon amplitudes, $F$, $G$, following Refs.\,\cite{Musatov:1999xp}.  This approach has achieved some phenomenological success (see {\it e.g.} \cite{Guidal:2013rya,Mezrag:2013mya}); but more flexible parametrisations enable a better fit to data \cite{Kumericki:2008di}.  Such fits play a valuable role in establishing the GPD framework;
However there is no known parameterization of GPDs relying on first principles only. Computing GPDs in a symmetry-preserving framework is a key ingredient for the {\it a priori} fulfillment of all GPD theoretical constraints.  This observation is highlighted by experience drawn from the simpler case of the pion's valence-quark PDF \cite{Chang:2014lva}.  
In \cite{Mezrag:2014tva,Mezrag:2014jka}, steps are made by following a different approach for the computation of hadron GPDs based on the example provided by the pion's valence-quark PDF. As sketched in \cite{Mezrag:2014jka} and elaborated futher in~\cite{Mezrag:2016hnp}, such a procedure only leaves with a reliable result near the so-called forward limit ($\xi=0$), within the DGLAP region~\cite{Dokshitzer:1977sg,Gribov:1972ri,Lipatov:1974qm,Altarelli:1977zs}. 

We will here briefly introduce the above-mentioned algebraic model for the pion's valence-quark GPD, obtain the overlap representation for the GPD for the same model and, preliminary, discuss a procedure, based on the Radon-transform technology for its parametrization, allowing for a reliable extension to the ERBL kinematical region~\cite{Efremov:1979qk,Lepage:1980fj}.

\section{The algebraic model for the pion's valence-quark GPD}
\label{sec-GPD}

The main features of the pion's valence-quark GPD has been recently sketched~\cite{Mezrag:2014jka} (although only within the DGLAP kinematical region) with the help of a simple algebraic model, anyhow able of a veracious pion's description, properly grounded in a faithful expression of their symmetries and their breaking patterns~\cite{Chang:2011vu,Bashir:2012fs,Qin:2014vya}, via the Dyson-Schwinger equations (DSEs). Such is the model we will exploit in the following, aiming at the comparison of its results with the corresponding overlap approach's, in DGLAP region, and at exploring the extension beyond via the Radon transform approach. 

Despite its complexity, the pion bound-state is still a $J=0$ system and hence there is only one GPD associated with a quark $q$ in the pion ($\pi^\pm$, $\pi^0$), which is defined by the matrix element
\begin{equation}
\label{eq-def-GPD-H-spinless-target}
H^q_{\pi}( x, \xi, t ) =  \int \frac{\mathrm{d}^4z}{4\pi} \, e^{i x P\cdot z}\,
\delta(n\cdot z) \, \delta^2(z_\perp) \,\langle\pi(P_+)| \bar{q}\left(-z/2\right)n\cdot \gamma \; \mathcal{W}(-z/2,z/2)
q\left(z/2\right) |\pi(P_-)\rangle,
\end{equation}
where $k$, $n$ are light-like four-vectors, satisfying $k^2=0=n^2$, $k\cdot n=1$; $z_\perp$ represents that two-component part of $z$ annihilated by both $k$, $n$; and $P_\pm = P \pm \Delta/2$ 
and $\mathcal{W}(-z/2,z/2)$ represents a Wilson line laid along a light-like path that joins the two listed vectors.
In Eq.\,\eqref{eq-def-GPD-H-spinless-target}, $\xi = -n\cdot \Delta/[2 n\cdot P]$ is the ``skewness'', $t=-\Delta^2$ is the momentum transfer, and $P^2 = t/4-m_\pi^2$, $P\cdot \Delta=0$. 
The GPD also depends on the resolving scale, $\zeta$.  Within the domain upon which perturbation theory is valid, evolution to another scale $\zeta^\prime$ is described by the ERBL equations \cite{Efremov:1979qk,Lepage:1980fj} for $|x|<\xi$ and the DGLAP equations \cite{Dokshitzer:1977sg,Gribov:1972ri,Lipatov:1974qm,Altarelli:1977zs} for $|x|>\xi$, where $\xi \geq 0$.

\begin{figure}
\centering
\begin{tabular}{cc}
\includegraphics[width=6.75cm,clip]{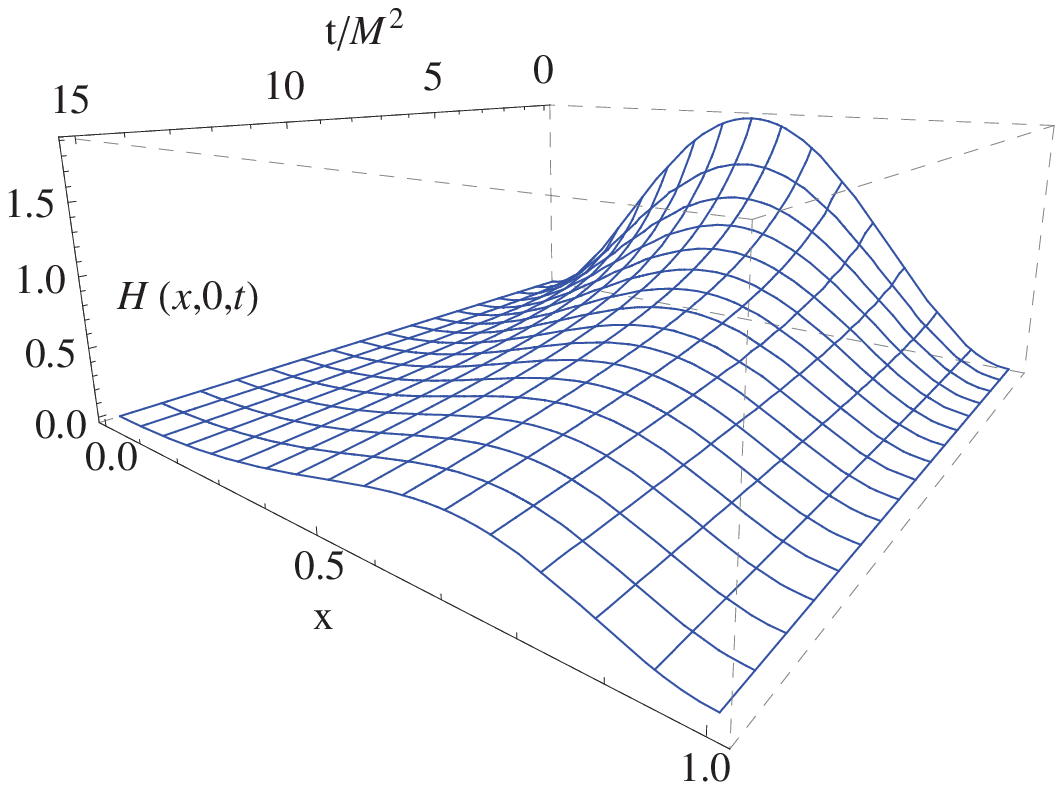} & 
\includegraphics[width=6.75cm,clip]{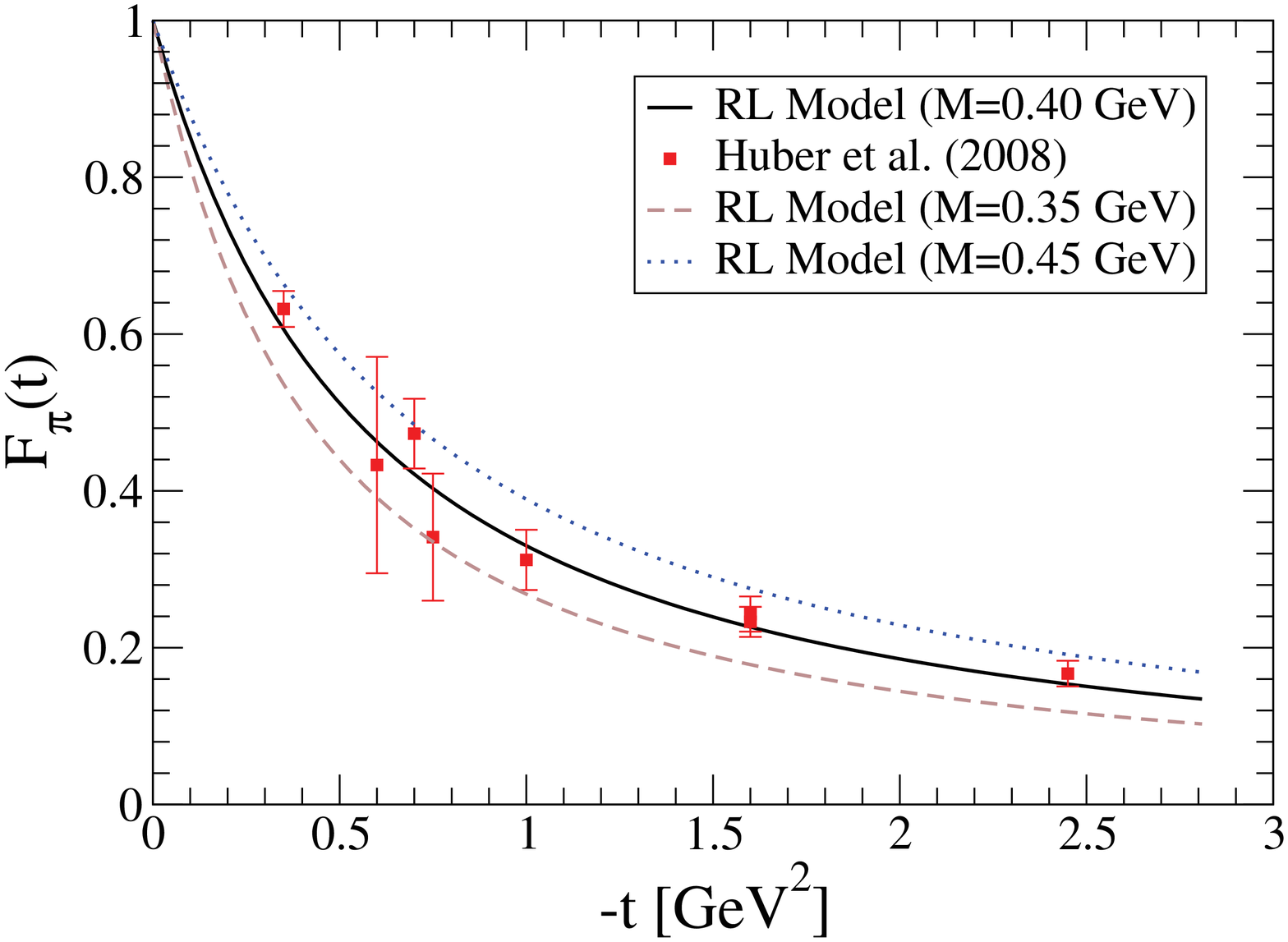}
\end{tabular}
\caption{\emph{Left panel}.- Pion valence dressed-quark GPD, $H_\pi^{\rm v}(x,0,-\Delta_\perp^2)$, defined by the addition of Eqs.\,(\ref{eq:TriangleDiagrams},\ref{HCorrection}), obtained as explained in~\cite{Mezrag:2014jka} and plotted as a function of $t/M^2=\Delta_\perp^2/M^2$, where $M$ is a mass scale for the dressed-quark in the algebraic model of \cite{Chang:2013pq} (result obtained at the model scale, $\zeta_H=0.51\,$GeV). \emph{Right panel}.-  Pion electromagnetic form factor obtained from the integtation over $x$ (sum rule) of $H_\pi^{\rm v}(x,0,-\Delta_\perp^2)$, resulting from Eqs.\,\eqref{eq:TriangleDiagrams}, \eqref{NakanishiASY} and associated definitions, in Ref.~\cite{Mezrag:2014jka}. The data are described in Ref.\,\protect\cite{Huber:2008id}. The most favourable comparison is obtained with $M=0.40\,$GeV in Eqs.\,\eqref{NakanishiASY} and the band shows results with $M=0.40\pm 0.05\,$GeV. 
}
\label{fig:GPDs}       
\end{figure}

As discussed at length in \cite{Mezrag:2014jka}, the valence-quark piece of the GPD expressed by \eqref{eq-def-GPD-H-spinless-target} can be first approximated by 
\begin{equation}\label{eq:TriangleDiagrams}
H_{\pi}^{\rm v}(x,\xi,t) = \frac 1 2 \;
 N_c \mathrm{tr}\rule{-0.5ex}{0ex}
 \int_{d\ell}\,\delta_n^{xP}(\ell)\,i{\Gamma}_\pi(\ell_+^{\rm R};-P_+ )\, 
S(\ell_+) \,in\cdot\Gamma(\ell_+,\ell_-) \, S(\ell_-) i\Gamma_\pi(\ell_-^{\rm R}; P_- )\, ,
\end{equation}
derived from the incomplete impulse-approximation for the so-called {\it handbag} diagram~\cite{Chang:2014lva}, and 
can be next corrected by the additional contribution
\begin{eqnarray}
H_\pi^{\rm C}(x,0,-\Delta_\perp^2) &=&
\frac{1}{2} N_c {\rm tr}\!\!\!
\int_{d\ell}\,\delta_n^{xP}(\ell)\left[
n\cdot \partial_{\ell_+^{\rm R}}\Gamma_\pi(\ell_+^{\rm R};-P_+)
S(\ell_P)\Gamma_\pi(\ell_-^{\rm R};P_-) \, S(\ell_-) \right. \nonumber \\ 
&& \left. + \Gamma_\pi(\ell_+^{\rm R};-P_+)S(\ell_P)
n\cdot \partial_{\ell_-^{\rm R}}\Gamma_\pi(\ell_-^{\rm R};P_-) S(\ell_-)
\right]\, , \label{HCorrection}
\end{eqnarray}
within the non-skewed kinematical region, for the transverse momentum $\Delta_\perp^2$. It is very worthwhile to emphasize here that, although \eqref{eq:TriangleDiagrams} paves the way for a fully general covariant computation of the pion's valence-quark GPD, the corrective term given by \eqref{HCorrection} can be hardly identified beyond DGLAP region and, even for the non-skewed case, some judicious modelling has been required for an appropriate description of the large-momentum transfer domain~\cite{Mezrag:2014jka}. In Eqs.~(\ref{eq:TriangleDiagrams},\ref{HCorrection}), $\int_{d\ell} := \int \frac{d^4\ell}{(2\pi)^4}$ is a translationally invariant regularisation of the integral; $\delta_n^{xP}(\ell):= \delta(n\cdot \ell - x n\cdot P)$; the trace is over spinor indices; $\eta\in[0,1]$, $\bar\eta=1-\eta$; $\ell_+^{\rm R}=\bar\eta\ell_+ +\eta\ell_P$,
$\ell_-^{\rm R}=\eta\ell_- +\bar\eta\ell_P$,
$\ell_\pm = \ell \pm \Delta/2$, $\ell_P=\ell-P$ (N.B.\ Owing to Poincar\'e covariance, no observable can legitimately depend on $\eta$; i.e., the definition of the relative momentum). In order to gain novel insights into pion structure, in Ref.~\cite{Mezrag:2014jka}, we used the algebraic model of \cite{Chang:2013pq}, 
\begin{subequations}
\label{NakanishiASY}
\begin{eqnarray}
\label{eq:sim1}
S(\ell) &=&[-i\gamma\cdot \ell+M]\Delta_M(\ell^2) \\
\rho_\nu(z) &=& \frac{1}{\sqrt{\pi}}\frac{\Gamma(\nu+3/2)}{\Gamma(\nu+1)}(1-z^2)^\nu \\
\label{eq:sim2}
n_\pi \Gamma_\pi(\ell^{\rm R}_\pm;\pm P) &=& i\gamma_5\int^1_{-1}dz\, \rho_\nu(z) \, \hat\Delta^\nu_M(\ell^2_{z\pm}) 
\label{NoF}
\end{eqnarray}
\end{subequations}
for the dressed-quark and pion elements in Eqs.~(\ref{eq:TriangleDiagrams},\ref{HCorrection});
where $\Delta_M(\ell^2)=1/(\ell^2+M^2)$, $M$ is a dressed-quark mass-scale; $\hat\Delta_M(\ell^2) = M^2 \Delta_M(\ell^2)$; $\ell_{z\pm}=\ell^{\rm R}_\pm + (z \pm 1) P/2$ and we work in the chiral limit ($P^2=0=\hat m$, where $\hat m$ is the current-quark mass); and $n_\pi$ is the Bethe-Salpeter amplitude's canonical normalisation constant. 
We then applied the algebra and approximations described in~\cite{Mezrag:2014tva,Mezrag:2014jka} to be finally left with the results displayed in Fig.~\ref{fig:GPDs} (left panel). Notably, the so-computed non-skewed GPD shows all the properties expected on the basis of the GPD overlap representation~\cite{Burkardt:2000za,Diehl:2000xz,Burkardt:2002hr,Diehl:2003ny}, discussed in \cite{Mezrag:2014jka}. Additionaly, integrated over $x$, it produces an estimate for the pion electromagnetic form factor which compares very well with experimental data (see right panel of Fig.~\ref{fig:GPDs}). 

Let us, however, focus on the forward limit for the non-skewed GPD obtained with eqs.(\ref{eq:TriangleDiagrams},\ref{HCorrection}), as described in \cite{Mezrag:2014jka}. It naturally reproduces the pion valence dressed-quark distribution function (PDF) found in \cite{Chang:2014lva}, observing, in particular, the symmetry under the exchange $x \to 1-x$ that owes to the two-body-problem nature for the pion's distribution at zero-momentum transfer when isospin-breaking contributions are not considered. Furthermore, as will be seen in the next section, the so-computed PDF can be compared with that obtained, for the same algebraic model, within the approach resulting from the GPD overlap representation~\cite{Burkardt:2000za,Diehl:2000xz,Burkardt:2002hr,Diehl:2003ny}.

\section{Overlap representation, Radon transform and skewed GPD}

As it has been previously discussed, in the aim of obtaining results beyond $\xi=0$, both in DGLAP and ERBL kinematical regions, a sensible extension of \eqref{HCorrection} correcting \eqref{eq:TriangleDiagrams} is far from being obvious. A different approach based on the representation of the pion GPD as  overlap of light-cone wave functions (LCWF), 
\begin{equation}\label{eq:overlap}
H_\pi^q(x,\xi,t)_{\xi \le x \le 1} = C^q \int d^2{\bf k}_\perp^2 \Psi^\ast\left(\frac{x-\xi}{1-\xi},{\bf k}_\perp+\frac{1-x}{1-\xi}\frac{\Delta_\perp}2 ; P_- \right)
\Psi\left(\frac{x+\xi}{1+\xi},{\bf k}_\perp-\frac{1-x}{1+\xi}\frac{\Delta_\perp}2 ; P_+ \right) \ ,
\end{equation}
can be followed, where $C^q$ is a normalization constant and the LCWF can be computed by integrating over $k^-$ the pion Bethe-Salpeter wave function $\chi_\pi(k,P)$ projected onto $\gamma^+ \gamma_5$
\begin{equation}\label{eq:LCWF}
\Psi\left(k^+,{\bf k}_\perp;P\right) = - \frac 1 {2\sqrt{3}} \int \frac{dk^-}{2\pi} 
\mbox{\rm Tr}\left[ \gamma^+ \gamma_5 \, \chi_\pi(k,P)\right] \ .
\end{equation}
The latter is only true within the DGLAP kinematical region, as only there the GPD results from the overlap of two $N$-body LCWF's (truncated to two-body in our pion's GPD case). Had the GPD overlap been lying within the ERBL region, it would have connected $N$- and $N$+2-body wave functions instead. One interesting remark in order here. The GPD can be represented by the Radon transform of DD's owing to fundamental properties it must fulfill, as Lorentz invariance. In addition, information from both DGLAP and ERBL regions is required in order to fully determine the DD representing the GPD. Consequently, by means of the Radon-transform representation, non-trivial connection between $N$- and $N$+2-body wave functions have to emerge from the fundamental symmetries of the physical world. Indeed, one can take advantage of the usually called one-component DD (1CDD) or, specifically, BMKS scheme~\cite{Belitsky:2000vk} (see also discussion in~\cite{Mezrag:2013mya} and references therein), and the GPD can be expressed as the Radon transform, $\mathcal{R}f$, of the distribution $f(\beta,\alpha,t)$,
\begin{eqnarray}
\label{eq:radon}
\frac{\sqrt{1+\xi^2}}{x} \, H(x,\xi,t) &=& \sqrt{1+\xi^2} \,  \int_{\Omega} d\beta \,  d\alpha \, 
 f(\beta,\alpha,t) \, \delta(x-\beta-\alpha \xi) \\ \label{eq:radon2}
&=& \mathcal{R}f(s,\varphi,t) = \int_{\Omega} 
d\beta \,  d\alpha \, f(\beta,\alpha,t) \,
\delta(s-\beta \cos{\varphi}-\alpha\sin{\varphi})\, ,
\end{eqnarray}
where the two DD variables $(\beta,\alpha)$ appear supported on a rotated square $\Omega \equiv \{ |\alpha|+|\beta|\leq 1\}$ and  \eqref{eq:radon2} makes explicit the geometrical interpretation of the Radon transform by the use of the polar coordinates which relate to standard GPD ones as $s=x\cos{\varphi}$ and $\xi=\tan{\varphi}$. Owing to its connection with the computerized tomography problem, mathematicians have paid enough attention to the Radon transform and provided us with a valuable tool for the characterization of solutions and their uniqueness under some smoothness and consistency conditions~\cite{Moutarde:2017}. 
Therefore, a natural way in order to make the extension for the GPD beyond DGLAP to ERBL region comes out from inverting \eqref{eq:radon} in order to get the corresponding DD from the DGLAP-constrained GPD. 

\begin{figure}
\centering
\sidecaption
\includegraphics[width=7cm,clip]{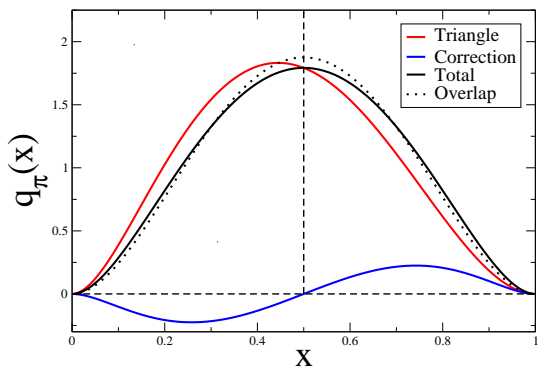}
\caption{Pion dressed-quark distribution function (PDF) obtained from the GPD in the forward limit ($t=0$, $\xi=0$); solid red and blue lines stand for the results from \eqref{eq:TriangleDiagrams} and \eqref{HCorrection}, respectively, and black solid one is for their addition; the dotted line corresponds to the overlap result given by \eqref{eq:PDF}, obtained from Eqs.~(\ref{eq:overlap},\ref{eq:LCWF}), for $\nu=1$.
}
\label{fig:PDF}       
\end{figure}

We will preliminary discuss how this can be acomplished in the next section, in particular by applying this program to the results  obtained with the previously described algebraic model. The first step is to use eq.\eqref{eq:LCWF} to get a Bethe-Salpeter wave function from \eqref{NakanishiASY}, which will be subsequently plugged into eq.\eqref{eq:overlap} and one will be finally left with the corresponding DGLAP overlap GPD. Thus, we obtain~\cite{Mezrag:2016hnp}
\begin{equation}\label{eq:GPDoverlap}
\left. H_{\pi^+}^{u}(x,\xi,0)\right|_{\rm DGLAP} \ = \ \mathcal{N}_\nu \frac{(1-x)^{2\nu} (x^2-\xi^2)^\nu}{(1-\xi^2)^{2\nu}} \ .
\end{equation}
If one takes now the forward limit ($t=0$, $\xi=0$) in \eqref{eq:GPDoverlap}, the result reads
\begin{equation}\label{eq:PDF}
q_\pi(x) \ = \ H^u_{\pi^+}(x,0,0) \ = \ \mathcal{N}_\nu \ x^{2\nu} (1-x)^{2\nu} \ ,
\end{equation}
where, in the case $\nu=1$, the normalization factor is $k_1=30$ and the result for the pion dressed-quark distribution function (PDF) has been proved to be numerically consistent with that obtained from Eqs.~(\ref{eq:TriangleDiagrams},\ref{HCorrection}), also in the forward limit. This can be seen in Fig.~\ref{fig:PDF}, where both results appear depicted and, to the eye, can be barely distinguishable from each other. In Ref.~\cite{Chang:2014lva}, the same result of \eqref{eq:PDF} with $\nu=1$ was introduced as an excellent and efficacious approximation to the pion's valence dressed-quark PDF, while also resulted, in \cite{Mezrag:2014jka}, from a heuristic LCWF implemented in \eqref{eq:overlap}. Here, it appears as the natural result from the algebraic model of Eqs.~\eqref{NakanishiASY}~\cite{Chang:2013pq} and the overlap representation of pion's GPD. 

\section{Beyond DGLAP} 

The final step of the computational program is then to make the inversion of the Radon transform in \eqref{eq:radon} with the DGLAP GPD given by eq.~\eqref{eq:GPDoverlap}. One will so obtain a DD representation for the GPD which determines it fully in both DGLAP and ERBL regions. It was anyhow discovered ~\cite{Teryaev:2001qm,Tiburzi:2004qr} that many parametrizations for DDs yield the same GPDs. These physically equivalent parametrizations can be mapped into each other by means of the so-called {\it gauge} transformations. In particular, there is a transformation that brings the DD to a particular representation,  $f_P(\beta,\alpha,t)$, due to Pobylitsa~\cite{Pobylitsa:2002ru} that, owing to a specific choice of the D-term, allows for the GPD to be recast as

\begin{eqnarray}
\label{eq:radonP}
H(x,\xi,t) \ = \ (1-x) \,  \int_{\Omega} d\beta \,  d\alpha \, 
 f_P(\beta,\alpha,t) \, \delta(x-\beta-\alpha \xi)  \ ,
 \end{eqnarray}
which will be particularly useful below. We will now specialize for the asymptotic value of $\nu=1$ in \eqref{eq:GPDoverlap} and will only consider here the case $t=0$. Then, in the following and for the sake of economy in the notation, $t$ will be dropped from the GPDs and DDs arguments. 

We have afforded the inversion of the Radon transform by properly discretizing the support domain $\Omega$ and so be left with a linear inversion problem in a large-dimension matrix space, $A x=y$, the dimension determined by our discretization choice, where the linear operator, $A$, expresses the action of the Radon transform over the discretized DD space. This is anyhow an {\it ill-posed} problem, at least in a mild sens, fom the mathematical point of view and much care needs to be taken in facing it numerically. In our case, the DD solution has been constrained by implementing two physically-grounded properties as (i) the parity in $\alpha$, $f(\beta,-\alpha)=f(\beta,\alpha)$; and, provided that we deal with a  valence-quark GPD, the kinematics requires that $f(\beta,\alpha)=0$ for any $\beta<0$. Both requirements work for any choice of the gauge and have been externally imposed. 
In particular, parity in $\alpha$ reduces the dimension of the matrix and requires a symmetrization of the discrete contributions coming from both sides of the $\alpha=0$ line. More details of the computation, in particular concerning the numerics for the Radon transform inversion, and a full kinematical analysis including the impact-parameter-dependent GPD  will be found in a forthcoming work~\cite{Chouika:2017}.  

\begin{figure}
\centering
\begin{tabular}{cc}
\includegraphics[width=6.95cm,clip]{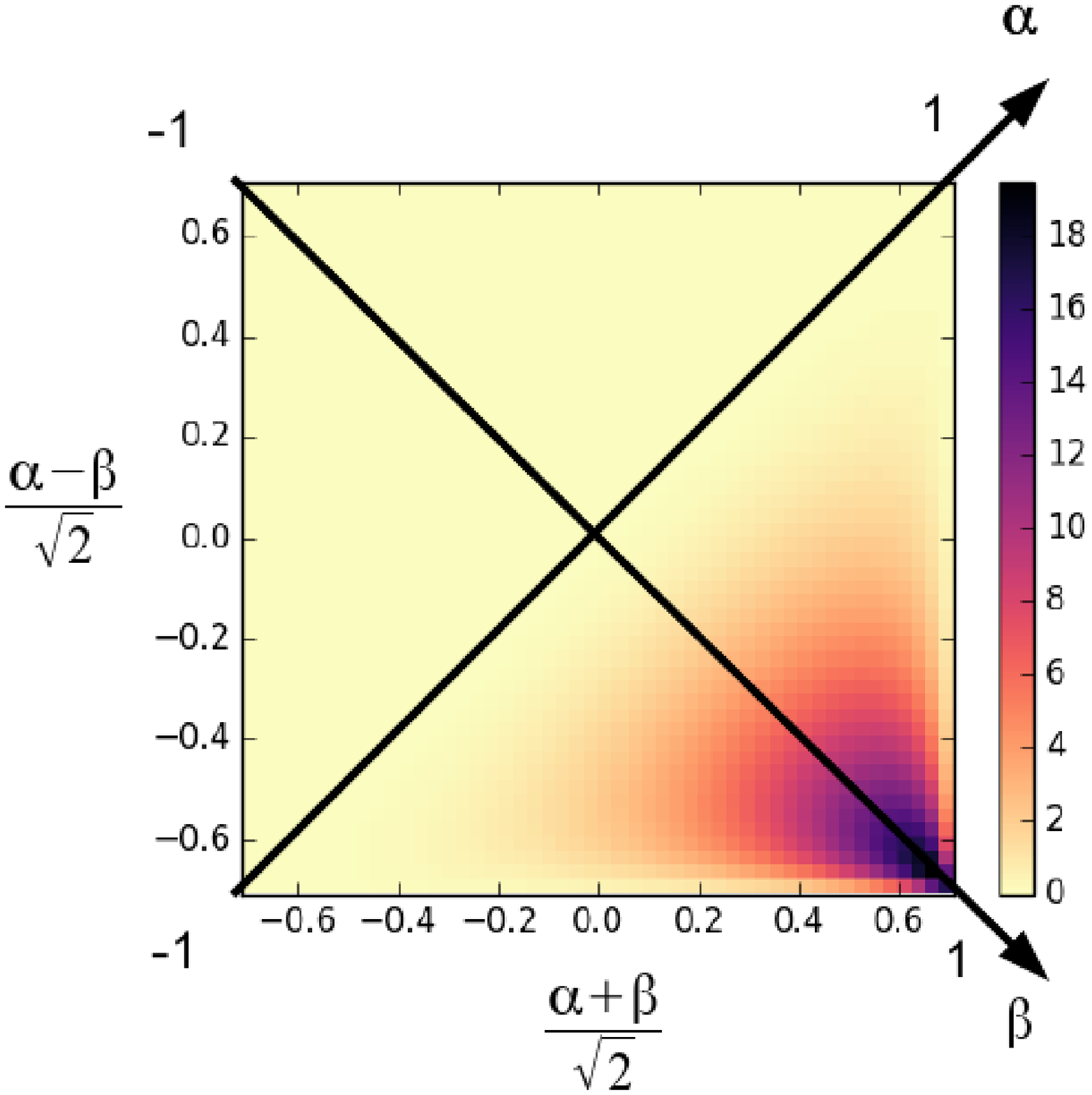} & 
\includegraphics[width=6.95cm,clip]{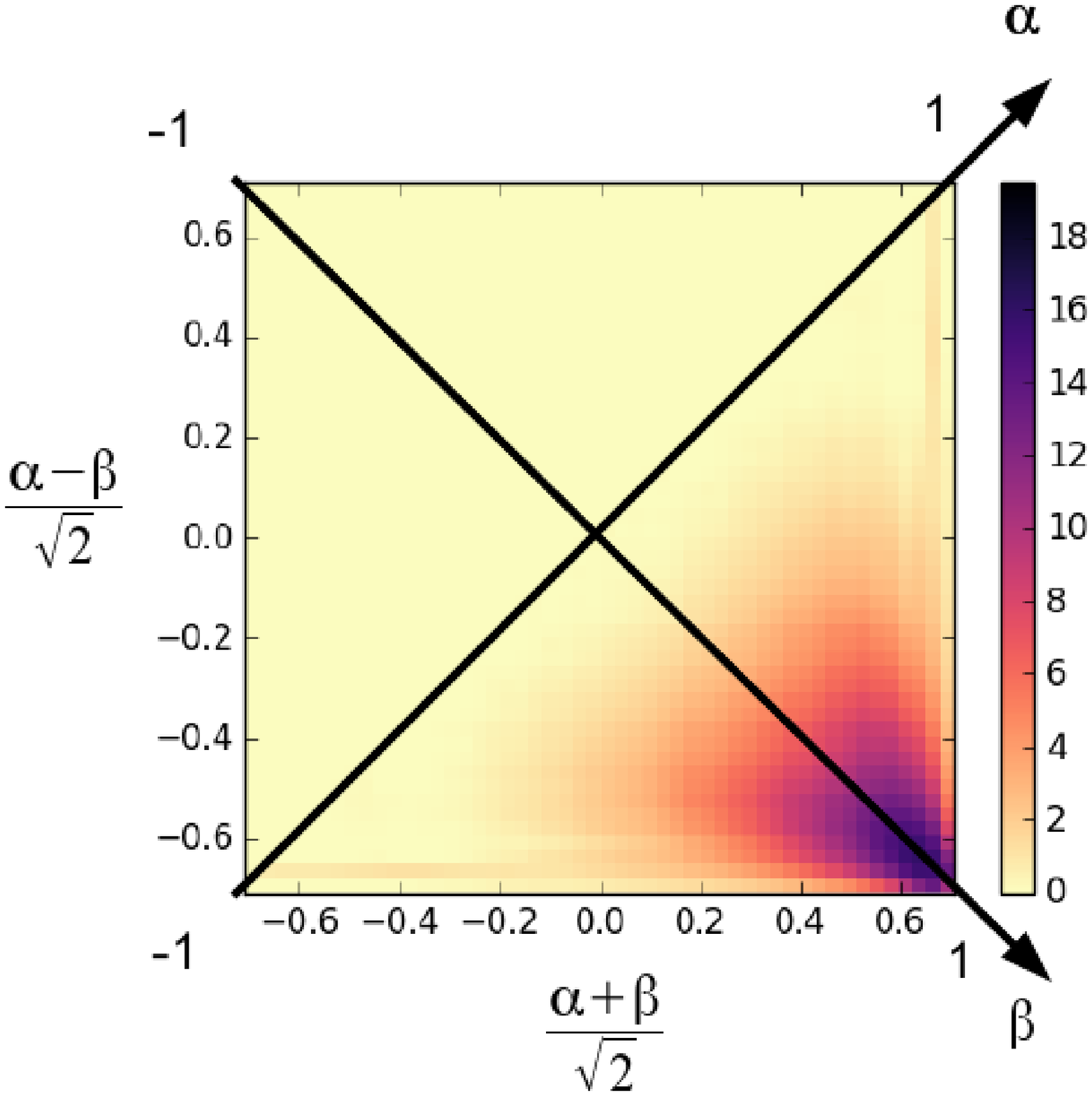}
\end{tabular}
\caption{Pobylitsa-gauge RDDA for $t=0$ obtained from the analytical expression, eq.~\eqref{eq:RDDA}, in the left panel; wich compares pretty well to the result of the inversed Radon-transform of its associated GPD in the DGLAP region, numerically obtained from \eqref{eq:radonP} as explained in the text, displayed in the right panel. The support for the parameters ($\alpha$,$\beta$) corresponds to a rotated square that we have rotated back by the choice of the two combinations, $(\alpha+\beta)/\sqrt{2}$ and 
$(\alpha -\beta)/\sqrt{2}$, for the horizontal and vertical axes, respectively. 
}
\label{fig:RDDAs}       
\end{figure}

Apart from toy examples, as a constant DD over the support domain where $\beta >0$, we have also first succesfully checked our numerics with a simplified version of the so-called Radyushkin Double Distribution Ansatz (RDDA)~\cite{Musatov:1999xp},
\begin{eqnarray}\label{eq:RDDA}
\left. \begin{array}{lc} 
\mbox{\rm Pobylitsa} ~~~& (1-\beta) \\
\mbox{\rm BMKS} ~~~& \beta 
\end{array} \right\} \, \times f_{RDDA}(\beta,\alpha) = \frac{\displaystyle\Gamma(N+3/2)}{\sqrt{\pi}\Gamma(N+1)} 
\frac{\left[(1-|\beta|)^2-\alpha^2\right]^N}{(1-|\beta|)^{2N+1}} \ q(\beta) \ ,
\end{eqnarray} 
where we specialized for $N=1$ and for a simple PDF resulting from $q(\beta)=\beta^2(1-\beta)^2$. The GPD associated to this RDDA can be analytically computed within the DGLAP region, and then numerically inverted and found to agree pretty well with the original DD given by \eqref{eq:RDDA}, as can be seen in fig.~\ref{fig:RDDAs}. 

\begin{figure}
\centering
\begin{tabular}{cc}
\includegraphics[width=6.95cm,clip]{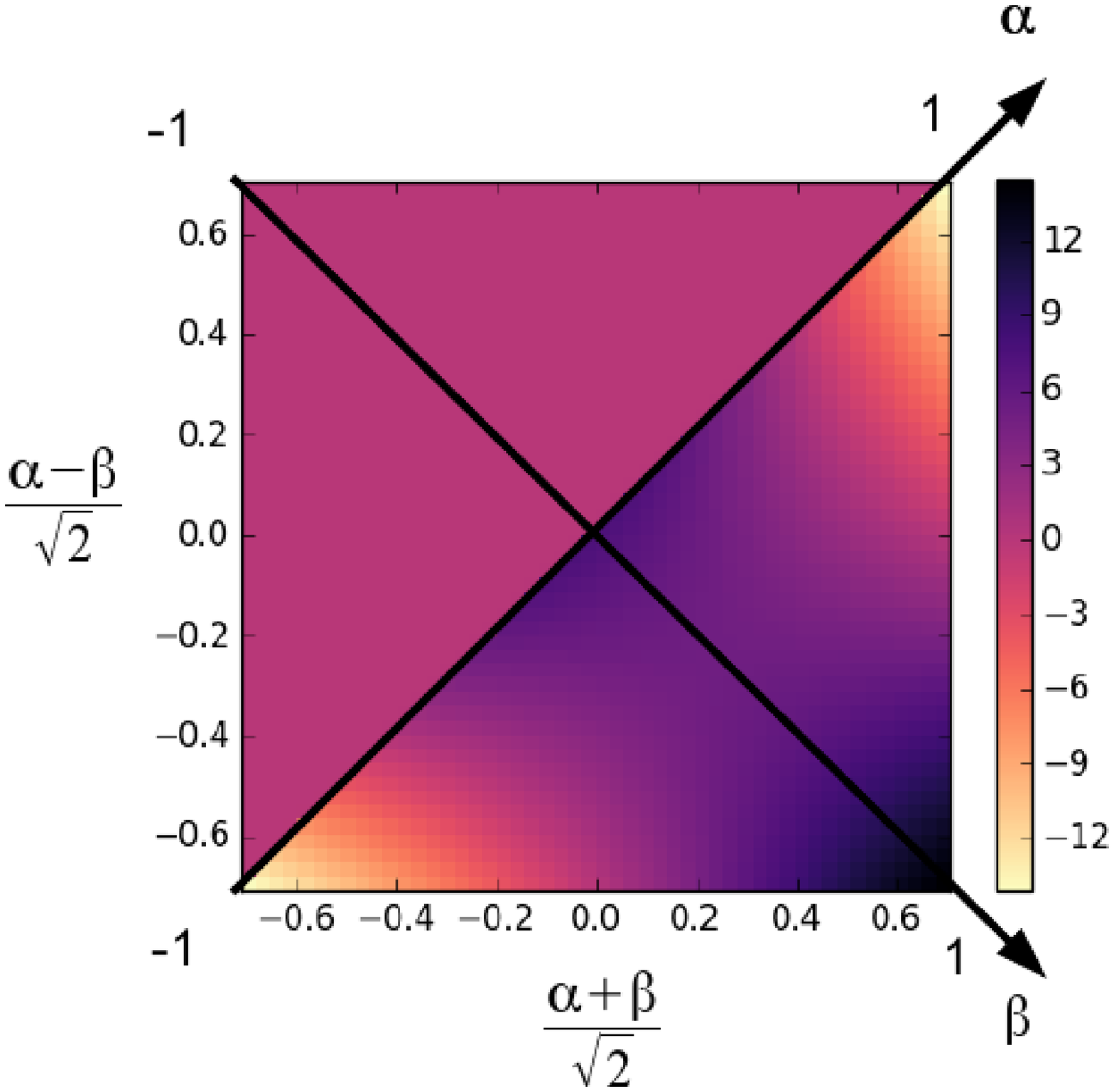} & 
\includegraphics[width=6.95cm,clip]{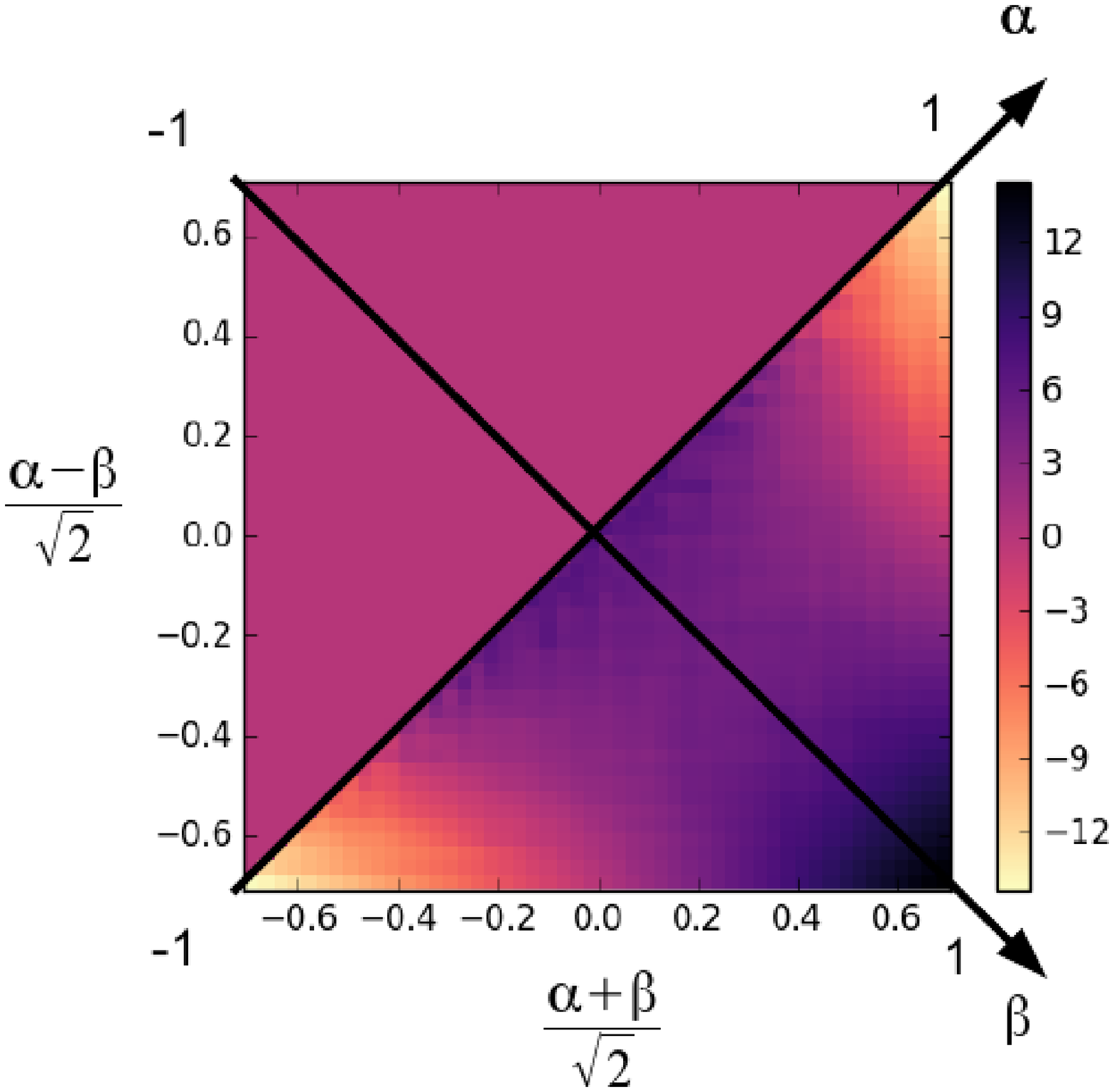}
\end{tabular}
\caption{Pion's valence dressed-quark overlap GPD at $t=0$ given by eq.~\eqref{eq:DDoverlapP} in the Pobylitsa representation, in the left panel; which also compares pretty well to the result of the inversed Radon-transform of its associated GPD in the DGLAP region, numerically obtained from \eqref{eq:radonP} as explained in the text, displayed in the right panel. 
}
\label{fig:overlaps}       
\end{figure}

Then, after succeeding in inverting the DGLAP GPD derived from the RDDA, we have a solid basis to rely on the following results obtained from applying the same numerical procedure to the overlap GPD given by \eqref{eq:GPDoverlap}. In that case, we found a much smoother solution by inverting the GPD expressed as a Radon transform representation in the Pobylitsa gauge, eq.~\eqref{eq:radonP}. Indeed, the numerical inversion yielded, in this case, a solution that can be easily accomodated in a polynomial in both, $\alpha$ and $\beta$, which can be suggested as an ansatz for the direct Radon transform and so get a solution to be compared with \eqref{eq:GPDoverlap}. Thus, we can conclude that 
\begin{equation}\label{eq:DDoverlapP}
f_P(\beta,\alpha) \ = \ \frac{30}4 \left( 1 - 3 \alpha^2 - 2 \beta + 3 \beta^2 \right) 
\end{equation}  
is the analytical Pobylitsa  DD for pion's valence-quark overlap GPD of the algebraic model given by \eqref{NakanishiASY} for the $t=0$ case (see fig.~\ref{fig:overlaps}). The mapping from Pobylitsa into BMKS gauge is then possible~\cite{Mueller:2014hsa} and we find

\begin{eqnarray}\label{eq:DDoverlap1CDD}
f(\beta,\alpha) &=& \frac{30}4 \left( \alpha^2 \left(3 - \frac{3}{(|\alpha|+\beta)^3}\right) + (3 \beta^2 -4 \beta) 
\left( \frac{1}{(|\alpha|+\beta)^4} - 1 \right) +  \frac{2}{(|\alpha| + \beta)^2} -2 \right. \nonumber \\
 &+& \left. (1-\alpha^2) \delta(\beta) \rule[0cm]{0cm}{0.5cm}\right) \ ,
\end{eqnarray}
which shows a singular behaviour in the $\beta=0$ line (this is precisely why the Pobylitsa gauge appears to offer a more friendly DD representation for the GPD concerning the inversion problem). 

Finally, once the Radon-transform inversion of \eqref{eq:GPDoverlap} has furnished us with any of the two DDs in eqs.~(\ref{eq:DDoverlapP},\ref{eq:DDoverlap1CDD}) for Pobylitsa and BMKS gauges (analytical expressions in this case, for $t=0$, but numerical results in general), we can apply either \eqref{eq:radonP} or \eqref{eq:radon} to obtain the Radon transform everywhere. 

\section{Conclusions}

We have briefly reported on the first steps recently made towards the computation of the pion's valence dressed-quark GPD, within the symmetry-preserving framework provided by DSEs. In addition,  we have also preliminarly introduced a reliable procedure, on the basis of the GPD representation as the overlap of LCWFs and Radon-transform technology, that can be fruitful exploited in order to extend previous results; in particular, those for the pion's GPD in DGLAP kinematical region and near the forward limit to non-zero skewness and to the ERBL domain. In addition, this procedure happens to be an open avenue for generic GPD modelling and, in the future, for the ambitious goal of computing GPDs from very first principles.

\begin{acknowledgement}

This work has been partially supported by Spanish ministry projects FPA-2014-53631-C2-2-P, French GDR 3034 PH-QCD ``Cromodynamique Quantique et Physique des Hadrons" and ANR-12-MONU-0008-01 ``PARTONS" and  U.S. Department of Energy, Office of Science, Office of Nuclear Physics, under contract no. DE-AC02-06CH11357. We thank for their very valuable remarks to  L.~Chang, C.D.~Roberts, F.~Sabati\'e and S.M.~Schmidt.

\end{acknowledgement}


\end{document}